\documentclass[aps,prl,reprint,superscriptaddress]{revtex4-1}

\usepackage{graphicx}
\usepackage{epsfig}
\usepackage{amsmath,amsthm}
\usepackage{textcomp}
\usepackage{color}
\usepackage{empheq}
\usepackage{multirow}

\begin{document}


\title{Local density of states in the superconductor $\kappa$-(BEDT-TTF)$_2$Cu[N(CN)$_2$]Br}


\author{Sandra Diehl}
\affiliation{Graduate School Materials Science in Mainz, Staudingerweg 9, 55128 Mainz, Germany}
\affiliation{Institut f\"ur Physik, Johannes Gutenberg-Universit\"at Mainz, Staudingerweg 7, 55128 Mainz, Germany}
\email{sandradiehl@uni-mainz.de}
\author{Torsten Methfessel}
\affiliation{Institut f\"ur Physik, Johannes Gutenberg-Universit\"at Mainz, Staudingerweg 7, 55128 Mainz, Germany}
\author{Jens M\"uller}
\author{Michael Lang}
\author{Michael Huth}
\affiliation{Physikalisches Institut, Goethe-Universit\"at Frankfurt, Max-von-Laue-Str. 1, 60438 Frankfurt am Main, Germany}
\author{Martin Jourdan}
\author{Hans-Joachim Elmers}
\affiliation{Institut f\"ur Physik, Johannes Gutenberg-Universit\"at Mainz, Staudingerweg 7, 55128 Mainz, Germany}


\date{\today}

\begin{abstract}
Low temperature scanning tunneling spectroscopy reveals the local density of states of the organic superconductor  $\kappa$-(BEDT-TTF)$_2$Cu[N(CN)$_2$]Br, that was cut in-situ in ultra-high vacuum perpendicular to the superconducting BEDT-TTF layers. The spectra confirm that superconductivity is confined to the conducting BEDT-TTF layers, while the Cu[N(CN)$_2$]Br anion layers are insulating. The density of states comprises a twofold superconducting gap, which is attributed to the two separated bands crossing the Fermi surface.
\end{abstract}


\maketitle

Superconductivity in non-metallic systems and in particular in organic materials has shown a variety of 
unconventional properties. The prevailing many-body interactions responsible for superconductivity
in low-dimensional electronic systems still prevent a full understanding of the nature of the superconducting pairing mechanism,
thus leading to vivid experimental and theoretical research~\cite{Mann2011}.  
Besides the symmetry of the order parameter that has been intensively studied in the past, 
less experimental effort has been devoted to the 
predicted consequences of multi-band superconductivity~\cite{Suhl1959} and 
only few firm confirmations of multiple superconducting energy gaps 
have been reported for MgB$_2$~\cite{Iavarone2002} 
and Ba$_{0.6}$K$_{0.4}$Fe$_2$As$_2$~\cite{Shan2011}, 
both cases representing superconductors with isotropic order parameter.
For a thorough understanding of multiband superconductivity it is important to clarify
whether  multigapped order parameters exist in strongly correlated anisotropic
systems, too.

Organic charge transfer (CT) salts such as  $\kappa$-(BEDT-TTF)$_2$Cu[N(CN)$_2$]Br
represent a model system~\cite{Toyota2007} for a 
strongly correlated superconductor with variable many-body interaction.
The superconducting properties of CT salts
 strongly deviate from  the behavior of BCS-like superconductors~\cite{Saito2007} but rather resemble the properties of 
high $T_c$-cuprates~\cite{McKenzie1997,Kanoda1997,Muller2002,Kagawa2005}.
The electronic properties of CT salts are closely related to their molecular structure thus leading to an electronic two-dimensional behavior~\cite{Lefebvre2000,Muller2002}.
The organic superconductor $\kappa$-(BEDT-TTF)$_2$Cu[N(CN)$_2$]Br ($\kappa$-Br) belongs to this type of CT salts and crystallizes in an orthorhombic unit cell~\cite{Kini1990,Kini1985}.
It is believed that the interesting electronic properties arise within the BEDT-TTF layer whereas the 
insulating anion layer controls the partial charge transfer and critically influences the electronic properties~\cite{Kanoda1997,Kagawa2005}.

A key property for a deeper understanding of electronic correlation effects is provided by the density of states (DOS) near the Fermi level~\cite{Ichimura2003,Ichimura2006,Nomura2009}.
Scanning tunneling spectroscopy~\cite{Arai2001} and photoemission spectroscopy~\cite{Downes2004} revealed the influence of the surface orientation for $\kappa$-(BEDT-TTF)$_2$Cu(NCS)$_2$.
However, only a few reports of real-space images of $\kappa$-(BEDT-TTF)$_2$X crystals have been published which were all performed at room temperature~\cite{Yoshimura1991,Fainchtein1992,Ishida2001}.
This precludes the simultaneous analysis of the structural and electronic properties because superconductivity in these salts only occurs at low temperatures.

We show STM images of in-situ cleaved surfaces of $\kappa$-Br at 5\,K. Simultaneously, we  measured differential conductivity spectra for anion and cation layers, confirming the two-dimensional superconductivity
within the BEDT-TTF layers.
We demonstrate the occurrence of
two energy gaps of different size simultaneously vanishing at $T_c$.
This observation provides a severe constriction to possible models for the superconducting paring mechanism.   

Single crystals of $\kappa$-Br were grown in an electrochemical crystallization process~\cite{Anzai1995}. All investigations were performed with a commercial low temperature scanning tunneling microscope under UHV-conditions with a base pressure of about $5\cdot10^{-11}$\,mbar. For obtaining a clean surface necessary for high resolution STM investigations we cleaved the crystal surface with a homebuilt in-situ cutter. The sample plate with the crystal is mounted into the already cooled down STM. The initial cooling rate of the crystals is of the order of 1K/min. We used an electrochemically etched tungsten tip flashed at about 2200\,K for all STM measurements. For the spectroscopic investigations we measured the voltage dependent tunneling current at a fixed tip position, averaged over at least 100 curves and differentiated them numerically to receive differential conductivity d$I$/d$V$ spectra. 
The measured spectra do not depend on the set point, excluding influences from local sample heating.

\begin{figure}[t]
\includegraphics[width=1.0\columnwidth]{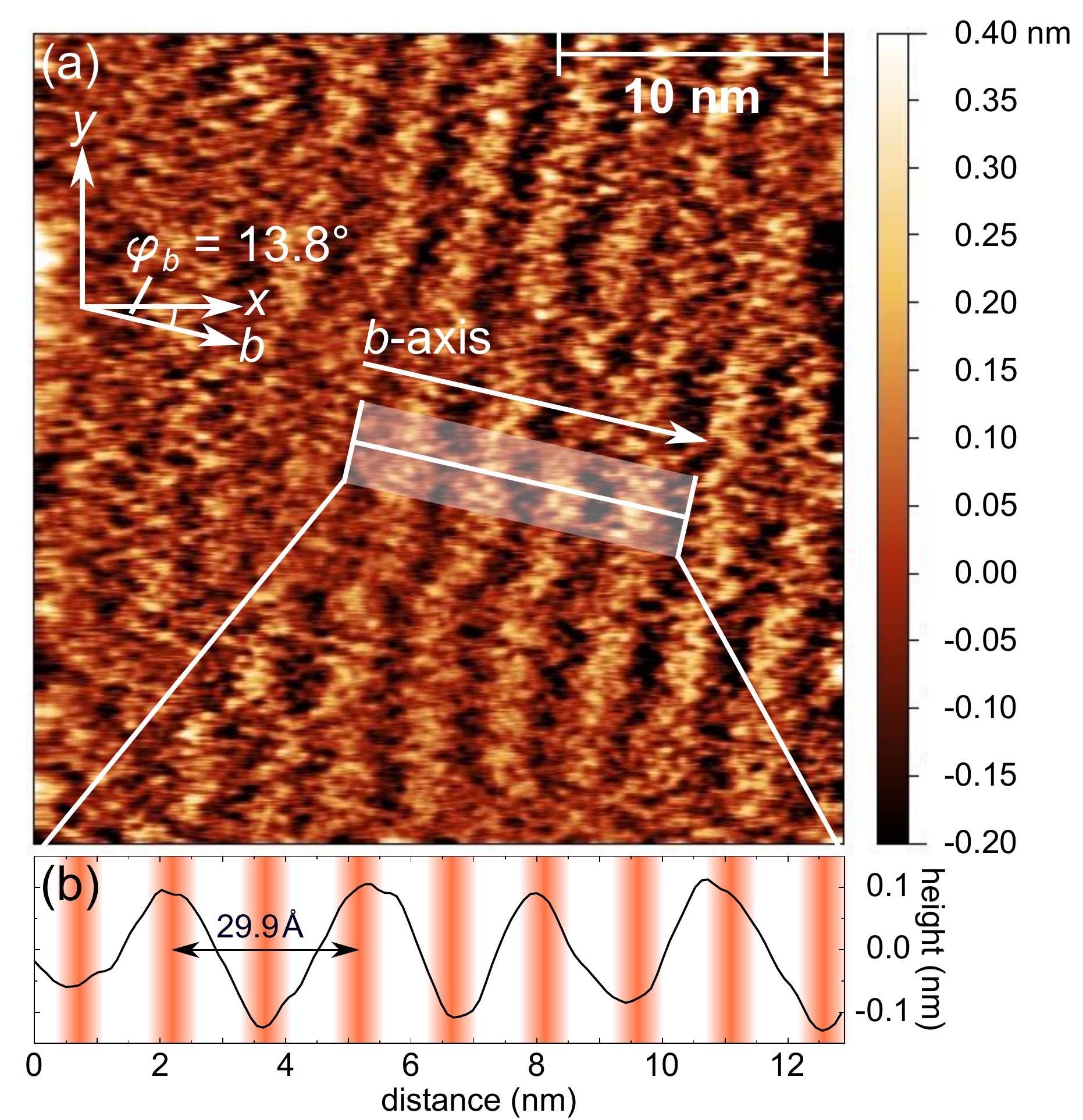}
\caption{\label{fig:30mnx30nm}(a) (30\,x\,30)\,nm$^2$ STM image of the crystal surface with tunneling direction parallel to the  layered structure of the $\kappa$-Br crystal revealing a stripe pattern ($T$\,=\,5\,K, $I$\,=\,60\,pA, $U_{\rm tip} = 30$\,mV). (b) The height profile along the marked area in (a) reveals an average stripe width of $29.9$.\,\AA. The red stripes indicate the position of the insulating layers.}
\end{figure}

\begin{figure}[t]
\includegraphics[width=1.0\columnwidth]{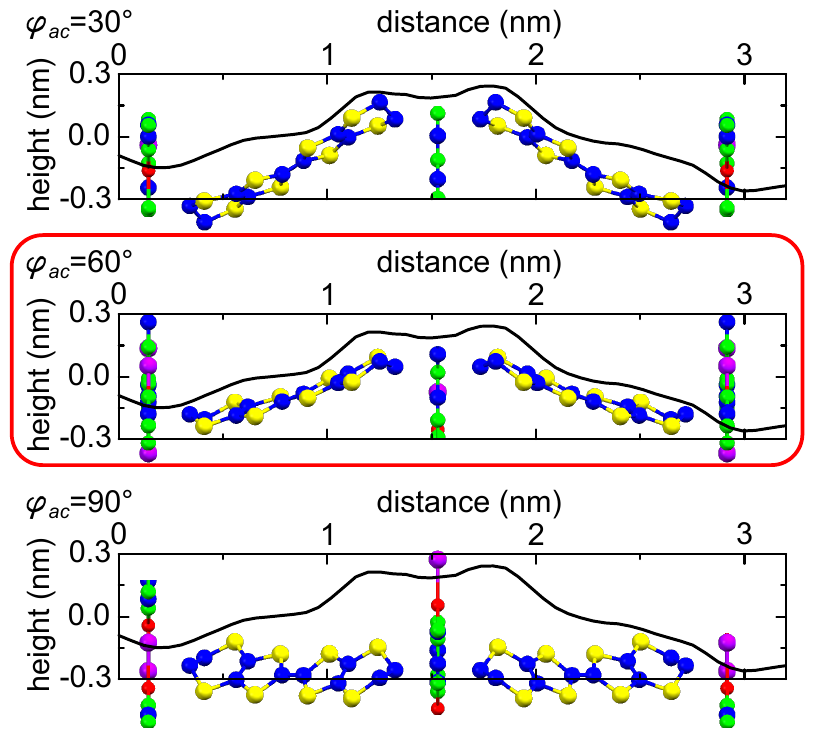}
\caption{\label{fig:heightprofile_single}Height profile along a single unit cell with the corresponding crystal orientation showing the best agreement for an angle for rotation about the b-axis of $\varphi_{ac} = 60\,^{\circ}$.}
\end{figure}

Fig.~\ref{fig:30mnx30nm}\,(a) shows a (30\,x\,30)\,nm$^2$ STM image of the $\kappa$-(BEDT-TTF)$_2$Cu[N(CN)$_2$]Br crystal cut perpendicular to the molecular layers.
The STM image reveals a stripe pattern with an average width of 29.9\,\AA\ [see Fig.~\ref{fig:30mnx30nm}\,(b)] which is in good agreement with the lattice constant of $\kappa$-Br in the $b$-direction. One stripe pattern consists of two BEDT-TTF and two anion layers. For a better illustration Fig.~\ref{fig:30mnx30nm}\,(b) shows the height profile along the white shaded area in (a) wherein the position of the anion layers is marked by red stripes. Hence the $b$-axis of the crystal is rotated by 13.8\,\textdegree\ with respect to the $x$-axis of the STM image. 
The height profile is mainly a mapping of the topographical surface profile, but is also influenced by the 
local density of states. Hence, the insulating anion-layers are expected to appear slightly lower (darker) than the conducting BEDT-TTF layers.
By analyzing the height profile along a single unit cell we determine the orientation of the cutting plane (see Fig.~\ref{fig:heightprofile_single}). Depending on the rotation about the $b$-axis (the axis perpendicular to the conducting layers)
the height corrugation caused by the alternating orientation of the BEDT-TTF molecules is more or less pronounced. The best agreement between height profile and crystal structure is found for a tilting angle $\varphi_{ac}=60$\textdegree\ about the $b$-axis with respect to the $a$-axis.

\begin{figure}[t]
\includegraphics[width=1.0\columnwidth]{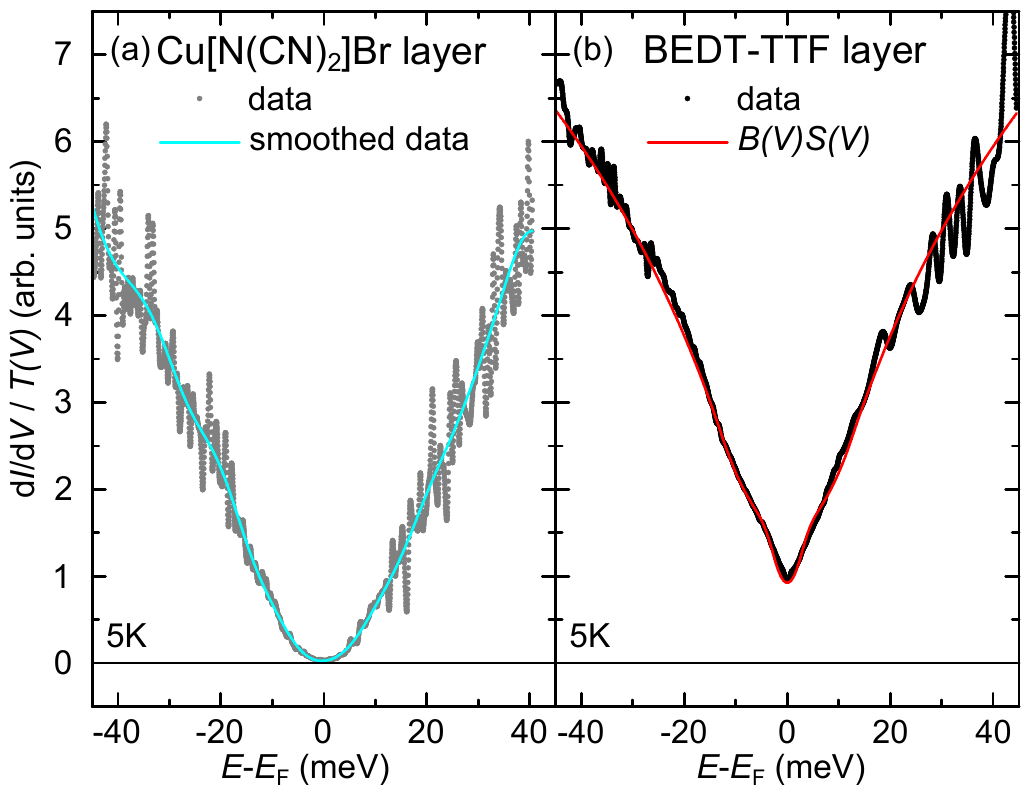}
\caption{\label{fig:spectra}Normalized differential conductivity spectra measured at different sample positions
with $-eU_{\rm tip}=eV=E-E_{\rm F}$.
(a) d$I$/d$V / T(V)$ measured at the Cu[N(CN)$_2$]Br layer. (b) d$I$/d$V / T(V)$ measured at the BEDT-TTF layer.}
\end{figure}

Next, the DOS of the anion and the BEDT-TTF layer was determined separately (see Fig.~\ref{fig:spectra}). 
Differential conductivity d$I$/d$V$ spectra were measured in the temperature range from 5\,K to 13\,K. The d$I$/d$V$ spectra were normalized to the tunneling transmission function $T(V)$, which is almost linear in the measured energy range and accounts for a small asymmetry observed in the original spectra~\cite{Ukraintsev1996}. Assuming a constant DOS for the tip material (d$I$/d$V$)/$T(V)$ is proportional to the DOS $D(V)$ of the sample.

The differential conductivity spectrum measured at the anion layer reveals an energy gap and shows no states at the Fermi edge [d$I$/d$V (E=E_F)\approx 0$] indicating an insulating behavior. This insulating behavior can be expected because the polymer layer comprises only strictly localized electronic states.

In contrast, the differential conductivity is non-zero  at $E=E_\mathrm{F}$ for the spectrum measured on the BEDT-TTF layer. The spectrum shows a $V$-shaped feature similar to the spectra measured with the STM tip perpendicular to the  layered structure~\cite{Diehl2014}. 
This observation confirms that the anion layer itself is insulating and the superconductivity occurs within the BEDT-TTF layers.

The $V$-shaped feature for $|eV|>15$~meV results from a logarithmic suppression of the density of states at the Fermi edge due to electronic disorder as discussed in Ref.~\cite{Diehl2014}. 
Temperature dependent measurements up to 13\,K show a temperature independent correction of the DOS which is described by~\cite{Shinaoka2009}
\begin{align}
B(V) = \exp \left[-\alpha\left(-\log \left| eV\right|\right)^d \right] 
\end{align}

where $d=2$ is the spatial dimension and $\alpha=0.3$ a non-universal constant~\cite{Diehl2014}.

\begin{figure}[t]
\includegraphics[width=1\columnwidth]{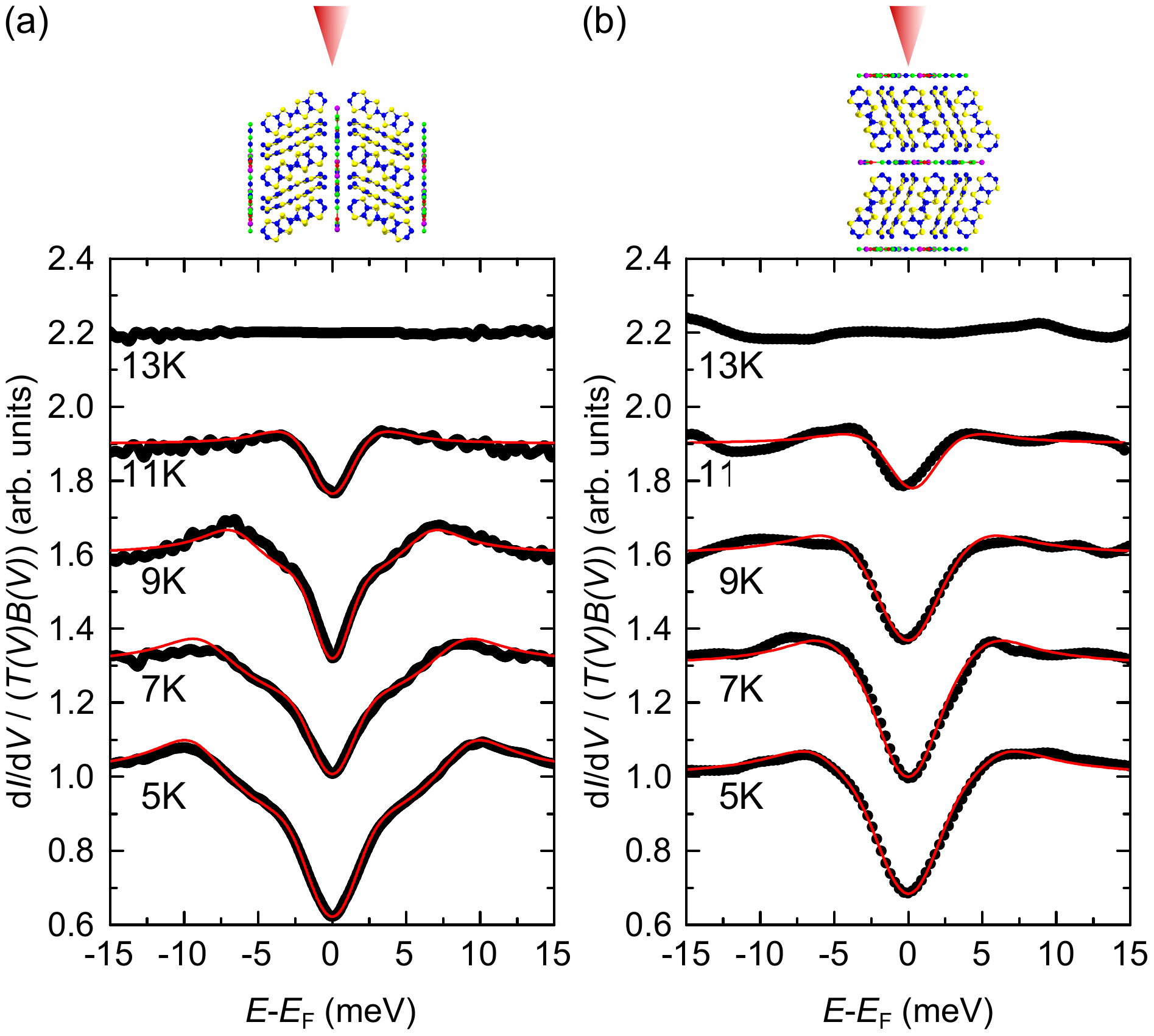}
\caption{\label{fig:sc_spectra}Superconducting energy gap at different temperatures measured parallel (a) and perpendicular to the layered crystal structure (b). The spectra have been measured at two different crystals. The measured data is fitted with eq.~\eqref{eq:Dynes} (red lines). The data shown in (b) is taken from Ref.~\cite{Diehl2014}.}
\end{figure}

\begin{figure}[t]
\includegraphics[width=1.0\columnwidth]{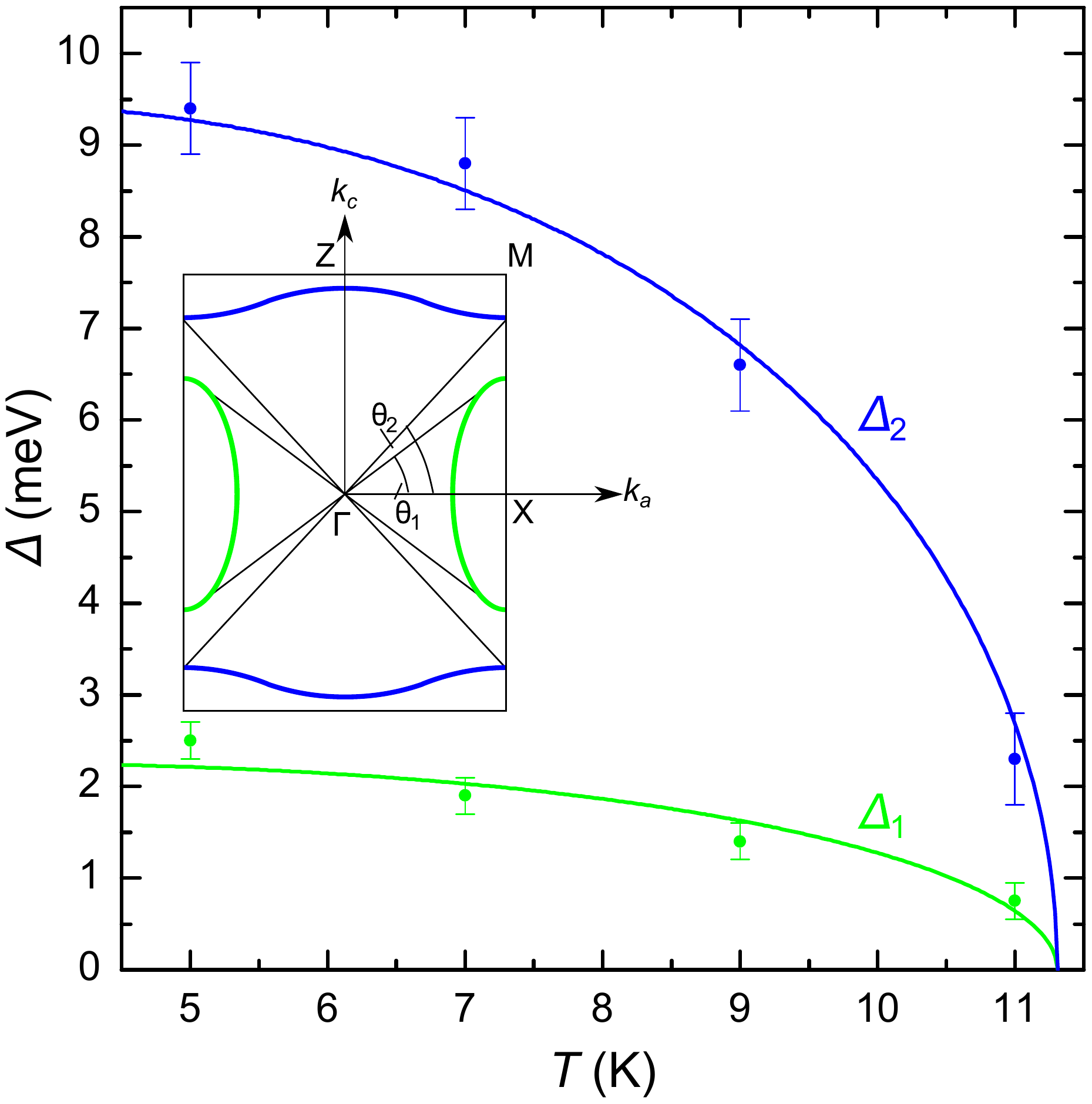}
\caption{\label{fig:deltatemp}Temperature dependence of the superconducting energy gaps. 
Full lines indicate fits to the BCS self-consistent equation.
The inset sketches the twofold Fermi surface of $\kappa$-Br} and the angular integration range.
\end{figure}

The density of states relating to the superconducting phase, S(V), is obtained via $S(V) = D(V)/B(V)$.
Focusing on the superconducting properties one obtains the spectra shown in Fig.~\ref{fig:sc_spectra}\,(a) revealing a clearly observable double gap structure.
$S(V)$ is fitted with a modified Dynes function~\cite{Dynes1978} 
assuming an angular dependence of the two order parameters. 
Additionally the imaginary part i$\mathit{\Gamma}$ of the energy respects a broadening of the measured gap: 
\begin{align}
d(\theta_1, \theta_2, V)&=\frac{c_1}{\theta_1} \int\limits_{0}^{\theta_1}\,\mathrm{d}\theta \frac{\mathrm{e}V + \mathrm{i} \Gamma}{\sqrt{(\mathrm{e}V + \mathrm{i}\Gamma)^2 - (\Delta_1 (\theta))^2}}\\ \nonumber
& +\frac{c_2}{\pi/2 - \theta_2} \int\limits_{\theta_2}^{\pi/2}\,\mathrm{d}\theta \frac{\mathrm{e}V + \mathrm{i} \Gamma}{\sqrt{(\mathrm{e}V + \mathrm{i}\Gamma)^2 - (\Delta_2 (\theta))^2}}
\end{align}
The integration boundaries $\theta_1$, $\theta_2$ are defined by the Fermi surface. The $k$-space integration leads to a suppression of contributions from possibly occurring nodes in the angular dependence of the gap $\mathit{\Delta}_i(\theta)$ such that an assumed $s$- and $d$-wave symmetry fits the experimental data almost equally well and leads to similar results for the gap energies.
To respect the thermal broadening this expression is convoluted with the Fermi function:
\begin{multline}
S\left(V\right) = \int\limits_{-\infty}^{\infty} \mathrm{d}E \frac{\partial f\left(E+\mathrm{e}V,T\right)}{\partial V} \cdot d(\theta_1,\theta_2,V)
\label{eq:Dynes}
\end{multline}
The obtained fitting values are listed in table~\ref{tab:delta}. 
The smaller gap contributes to about 80~\% to the measured gap ($c_1 \approx 0.8$)
whereas the wide gap is less pronounced ($c_2 \approx 0.2$). Both energy gaps decrease with increasing temperature and vanish above the critical temperature.
The temperature dependence can be described by
the BCS self-consistent equation resulting in values of $\Delta_1 (0\,\mathrm{K}) = (2.27 \pm 0.10)$\,meV, $\Delta_2 (0\,\mathrm{K}) = (9.51 \pm 0.17)$\,meV and $T_c = 11.3 \pm 0.2$\,meV for the perpendicular geometry
(see Fig.~\ref{fig:deltatemp}).

The extrapolated zero temperature energy gap $\mathit{\Delta}_1$(0\,K) of 2.27\,meV leads to a ratio of $2 \mathit{\Delta}_1(0\,\mathrm{K}) /(k_\mathrm{B} T_c )  = 4.66 \pm 0.22$ indicating  a strong coupling.
The large values of the broadening parameter $\mathit{\Gamma}$ may result from a disorder-induced broadening as discussed in Ref.~\cite{Diehl2014}.
Spectra measured parallel to the conducting layers differ from those measured perpendicular to the BEDT-TTF layers [see Fig.~\ref{fig:sc_spectra}\,(b)].
For the latter the gap energies differ by a factor of two 
instead of a factor of three making it harder to detect. As a consequence, assuming a single-gap state can still yield reasonable fits~\cite{Diehl2014}.
We attribute the observed differences to surface effects, e.g. structural relaxation, broken symmetry and decreased charge transfer because of missing layers, influencing the local density of states measured within the topmost molecular layer.
The surface effects will rather occur in the case of BEDT-TTF layers  lying parallel to the surface because the broken symmetry may easily lead to a relaxation of the topmost layers along the surface normal. 
In the case of the parallel orientation such a relaxation of the anion-cation layer distance would lead to macroscopic changes of the crystal shape and hence are suppressed. Therefore, we assume that the spectra measured parallel to the conducting layers [Fig.~\ref{fig:sc_spectra}\,(a)] present the bulk properties of the $\kappa$-Br crystals more accurately.  In contrast, the spectra measured in the perpendicular orientation are critically influenced by surface states which seems to lead to a reduced second gap.
\begin{table}[t]
\caption{Fit parameters from the fit with Eq.~\eqref{eq:Dynes}.}
\label{tab:delta}
\begin{tabular}{c c|c|c|c|c|c}
& T (K)	&	$\mathit{\Delta}_1$ (meV)	&	$\mathit{\Delta}_2$ (meV)	&	$\mathit{\Gamma}$ (meV) 	& c$_1$ & c$_2$		\\ \hline
\multirow{4}{*}{\rotatebox{90}{parallel}}
& 5	 	&	2.5(2)				&	9.4(5)					&	1.4(1)		& 0.8(1)	& 0.2(1)		\\
& 7	 	& 	1.9(2) 				&	8.8(5)					&	1.1(1)		& 0.8(1)	& 0.2(1)		\\
& 9	 	& 	1.4(2)				&	6.6(5)					&	0.3(1)		& 0.9(1)	& 0.1(1)		\\
& 11	& 	0.8(2)				&	2.3(5)					&	0.1(1)		& 0.8(1)	& 0.2(1)		\\ \hline
\multirow{4}{*}{\rotatebox{90}{perp.}}
& 5	 	&	2.8(2)				&	6.4(5)				&	1.8(1)		& 0.8(1)	& 0.2(1)		\\
& 7	 	& 	2.4(2) 				&	5.4(5)				&	1.5(1)		& 0.7(1)	& 0.3(1)		\\
& 9	 	& 	2.1(2)				&	4.8(5)				&	1.4(1)		& 0.8(1)	& 0.2(1)		\\
& 11	& 	1.4(2)				&	3.2(5)				&	0.8(1)		& 0.9(1)	& 0.1(1)		\\ \hline
\end{tabular}
\end{table}

For the tunneling direction parallel to the conducting layers
the presence of two gaps is obvious. For the tunneling direction perpendicular to the conducting layers
the two gaps nearly merge to a single gap and become obvious only
from the improvement of the numerical fit,
explaining 
previous data interpretation~\cite{Diehl2014,Ichimura2006}.

While tunneling spectroscopy confirms two-band superconductivity for
MgB$_2$~\cite{Iavarone2002} and Ba$_{0.6}$K$_{0.4}$Fe$_{2}$As$_{2}$~\cite{Shan2011}
with strong preference for $s$-wave order parameter,
it was assumed that in the case of organic superconductors Hubbard-type models can be reduced
to single-band models.
Only recently ab initio two-dimensional multiband low-energy models
indicated that
the minimal models to describe low-energy phenomena of
the organic compounds are multiband models~\cite{Nakamura2012}.

An explanation for a two-gap structure has been proposed in Ref.~\cite{Suhl1959}
considering two electronic bands contributing to the order parameter.
Two different bands forming the Fermi surface are indeed present for $\kappa$-Br crystals
~\cite{Oshima1988,Mielke1997}.
Two overlapping bands at the Fermi level with different occupation numbers will lead to two energy gaps of different size 
if interband scattering prevails over intraband scattering.
In this case the two energy gaps are not attributed to single electron bands
but rather represent the result of binding and antibinding hybridization of 
the twofold superconducting ground state.
Interband scattering requires a transfer of finite momentum. 
On this basis existing theories to the coupling mechanism may have to be critically re-assessed with regard to their compatibility with a finite-momentum inter-band coupling  scenario.

Please note that although finding a two-gap structure
in tunneling measurements
provides confirmation of a two-band system, the
anisotropic single-band scenario cannot be completely
excluded~\cite{Zehetmayer2013}.

To summarize, we 
have studied low-temperature scanning tunneling microscopy and spectroscopy on
 $\kappa$-(BEDT-TTF)$_2$Cu[N(CN)$_2$]Br with the tunneling direction
parallel and perpendicular to the layered structure.
The measured spectra confirm the restriction of conductance to the BEDT-TTF layers, while the anion layers are insulating.
Below the critical temperature the density of states comprises two 
superconducting energy gaps of different size.
The extrapolated ground state magnitude of the smaller gap is  2.3~meV.
The corresponding value of the larger gap is 9.4~meV for tunneling parallel
and 6.5~meV for tunneling perpendicular to the BEDT layers,
indicating the influence of surface reconstruction.
The observed multigap features provide strong evidence for 
multiband superconductivity in the title compound.

\begin{acknowledgments}
We thank the Deutsche Forschungsgemeinschaft (SFB/TR 49) and the Graduate School Materials Science in Mainz for financial support.
\end{acknowledgments}

\end{document}